\documentclass[authorversion,manuscript,screen,acmtog]{acmart}

\setcopyright{none}
\copyrightyear{2023}
\acmYear{2023}

\usepackage[ruled, vlined]{algorithm2e}
\usepackage{wrapfig}

\begin{document}

\title{The Influence of Variable Frame Timing on First-Person Gaming}

\author{Devi Klein}
\email{devik@nvidia.com}
\affiliation{%
  \institution{NVIDIA}
  \country{USA}
}
\affiliation{%
  \institution{UC Santa Barbara}
  \country{USA}
}
\author{Josef Spjut}
\orcid{0000-0001-5483-7867}
\email{jspjut@nvidia.com}
\author{Ben Boudaoud}
\email{bboudaoud@nvidia.com}
\author{Joohwan Kim}
\email{sckim@nvidia.com}
\affiliation{%
  \institution{NVIDIA}
  \country{USA}
}
\renewcommand{\shortauthors}{Klein et al.}

\renewcommand{\shorttitle}{Influence of Variable Frame Timing}

\begin{abstract}
Variable frame timing (VFT), or changes in the time intervals between discrete frame images displayed to users, deviates from our traditional conceptualization of frame rate in which all frame times are equal.
With the advent of variable refresh rate (VRR) monitor technologies, gamers experience VFT at the display.
VRR, coupled with increased display refresh rates and high-end hardware, enables smoother variation of frame presentation sequences.
We assess the effects of VFT on the perception of smoothness (experiment 1) and performance (experiment 2) in first-person shooter (FPS) gameplay by introducing frequent but relatively small (4-12 ms) variations in frame time around typical refresh rates (30-240 Hz).  
Our results indicate that VFT impacts the perception of smoothness. 
However, the results from experiment 2 do not indicate differences in FPS task performance (i.e., completion time) between variable and constant frame time sequences ranked equally smooth in experiment 1.
\end{abstract}

\ccsdesc[500]{Human-centered computing~Pointing devices}
\ccsdesc[500]{Human-centered computing~User studies}
\ccsdesc[500]{Applied computing~Computer games}

\keywords{frame rate, frame time, first-person targeting, first-person games, perception, performance}

\begin{teaserfigure}
    \centering
    \includegraphics[width=0.77\textwidth]{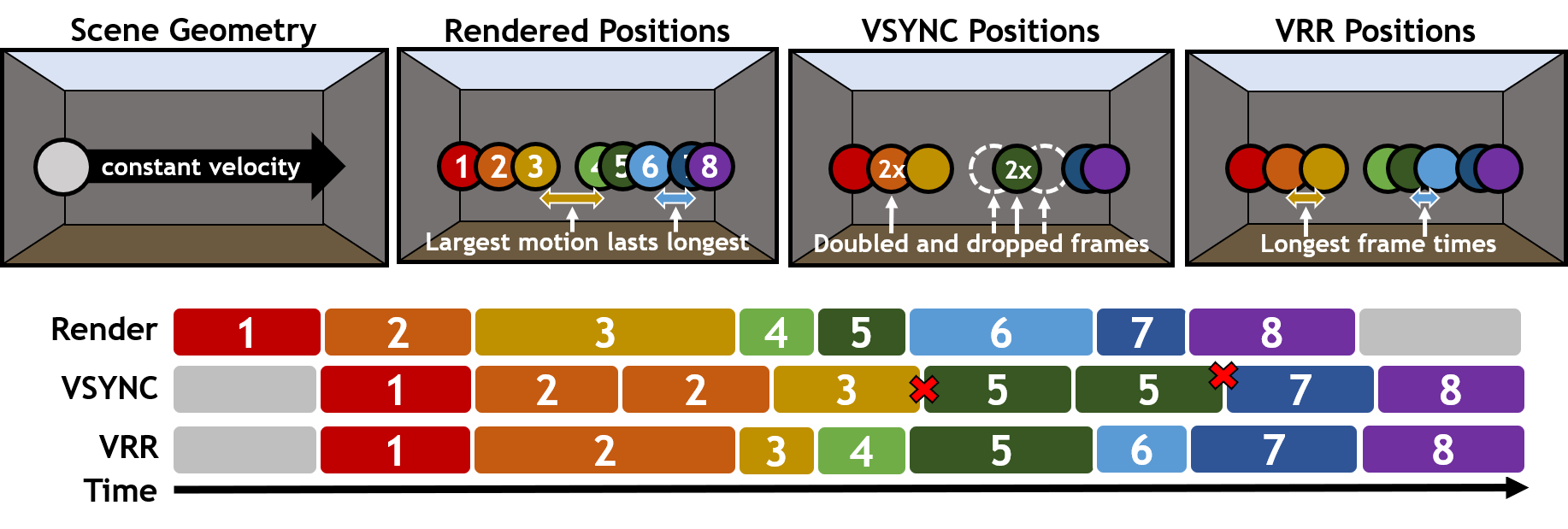}
    \includegraphics[width=0.22\textwidth]{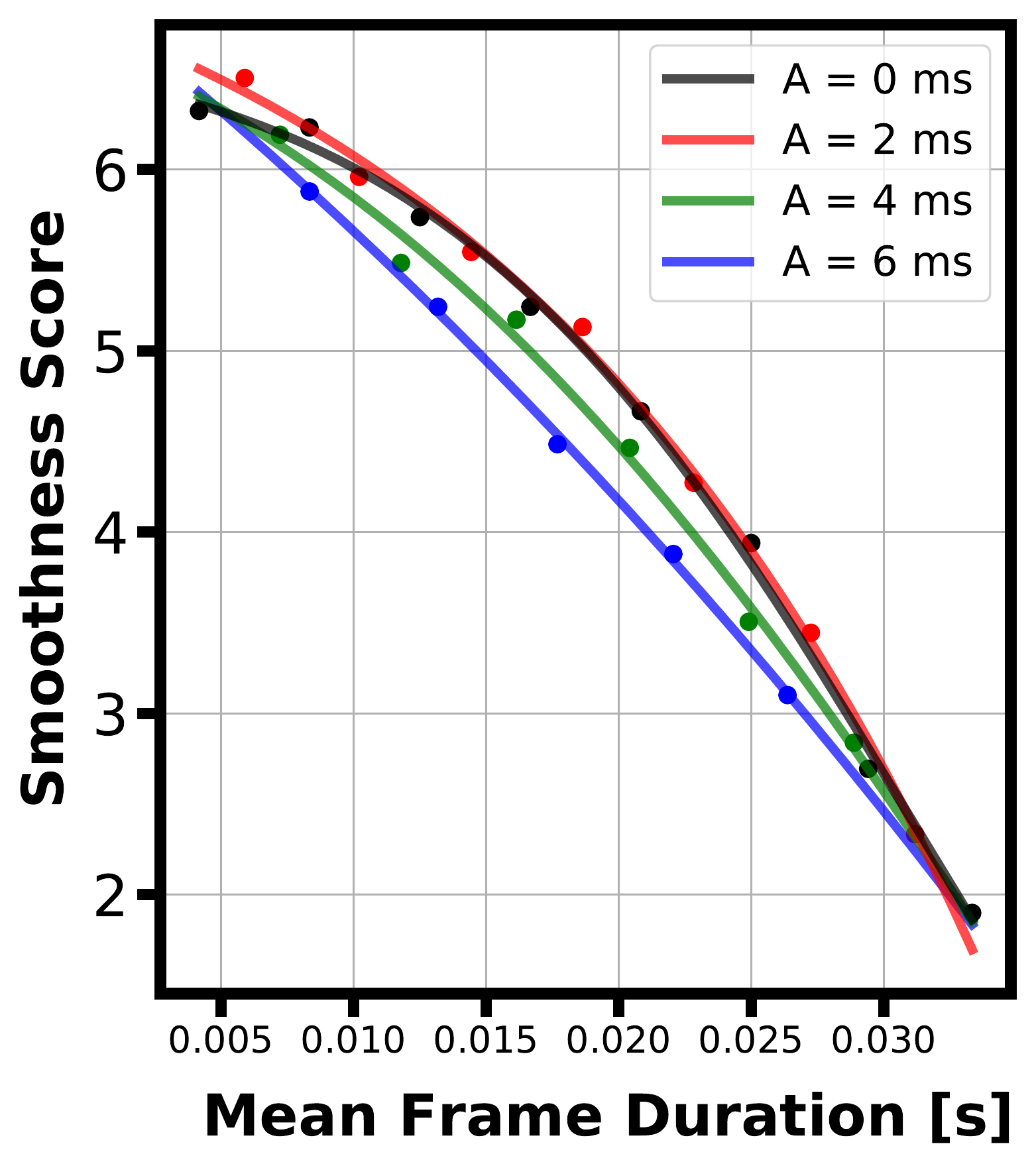}
    \caption{(Top Left) A demonstration of the visual effects of variable frame timing. (Bottom Left) A demonstration of how variable render/simulation workload timing impacts VSYNC and variable refresh rate (VRR) differently. 
    (Right) A plot of a user rating of "smoothness" under varying mean frame time and VFT amplitude (2-6 ms). Generally smoothness decreases with increased frame time, as expected, with additional degredations of smoothness coming from VFT.
    }
    \label{fig:teaser}
\end{teaserfigure}

\maketitle

\section{Introduction}
Our visual perception arises from a continuous-to-continuous mapping between the light reflected off objects in our 3D environment and entering our eye and the electrochemical transduction process occurring at the back of the retina \cite{rodieck1998first}. 
With the introduction of digital images, high refresh rate displays, and the quantization of light into discrete bits of information, modern technologies for visualization attempt to induce the spatio-temporal dynamics of the visual phototransduction process that drives real-world perception.
Despite the advances in emerging visualization technologies, the inherent discrete-to-continuous mapping of information can cause undesirable perceptual effects.

Take for example the prevalent problem of \emph{motion blur} on sample-and-hold-type displays such as liquid crystal displays (LCDs) \cite{10.1117/12.811757}. 
Assume a viewer who is visually tracking a moving object on an LCD operating at 60 Hz.
Because each LCD pixel ideally\footnote{LCD hold time and transition time are not necessarily a function of the frame period. 
Some transitions are longer than a frame period; others are short enough to meet this ideal, but the reality falls short in most LCDs.} holds its emission for the duration of one frame time (approximately 16 ms for a 60 Hz display), its optical image can smear across multiple retinal cells while the viewer's gaze continues to move to track the object.
Consequently, the viewer receives a blurry impression of the moving objects. Display artifacts such as motion blur occur when the discretized information cannot faithfully approximate the real-world experience for our visual system.

Prior work has focused on the visibility of motion artifacts for displays with constant frame time (CFT) (Sec.~\ref{sec:motion_artifacts}). 
CFT, however, is more of an idealized phenomenon in gaming due to variations in render time, input sampling, or network delays. 
Instead, high-end modern display technologies (e.g., variable refresh rate displays or VRR) present users with a sequence of frames that can persist on screen for different periods, a phenomenon we denote as variable frame timing (VFT). 
That is, one frame could last on screen for ~16ms (60Hz), and the next frame could last on screen for ~20 ms (50Hz).

Given the prevalence of VFT in modern-day gaming, and the relatively little empirical work investigating its influence on both the perception of smoothness and the effect on performance in an FPS-style targeting task, our paper provides the following key contributions:

\begin{enumerate}
    \item An introduction to variable frame timing (VFT). What is it? Why does it occur?
    \item An exploration of the effects of VFT on human perception (experiment 1) and performance (experiment 2) across a broad range of baseline frame rates. 
    \item Establishing a relationship between the perception of VFT and constant frame times (CFT).
    \item Provide suggestions for gamers who want to utilize VFT in both professional and leisure settings and for researchers who want to explore VFT in contexts outside of gaming.
\end{enumerate}

\section{Background}
The following sections provide a short primer on VFT and frame timing. The discussion points and background information underscore the importance of studying VFT in the wild because it is becoming more prevalent as users modernize their hardware and software to state-of-the-art technology. We follow this discussion with a brief overview of the perceptual quality and performance metrics used in experiments 1 and 2.

\subsection{Where does Variable Frame Timing (VFT) come from?}
There are many sources of VFT, including changes in game geometry or shading workloads, power state and cooling of PC hardware, operating system preemption and scheduling, network communication with a server, and even human interaction with the game.
While these sources interact in different ways, the result is that game frames are completed after the combination of a variety of different processing times, with each of these times introducing a possible source of variability.
In general, game frame time variability is relatively small compared to the total frame time, though this varies from game to game.
Game developers often target a single constant display output rate and conduct significance testing and optimization to ensure that most frames finish within the fixed budget.
This constant frame time target avoids bothersome interactions with display synchronization (i.e., VSync) that can amplify VFT effects when a game overshoots the target refresh period of the display (e.g., draws a frame for more than 33.3 ms for a 30 Hz target).


\subsection{Frame Rate, Frame Time, and Refresh Rate}
Many gamers and researchers use the terms \emph{frame rate} and \emph{frame time} interchangeably \cite{claypool2006effects}, but it is important to recognize their relationship and distinct implied meaning.
Frame \emph{rate} is the number of frames delivered over some timeframe, typically reported in Hertz (Hz) or the number of frames per second.
Frame \emph{time} is the time that elapses between a pair of frames during which the first frame is displayed, often the time it took to complete the rendering and display of the second frame in that pair.
Assuming frames are completed on a regular schedule, the frame rate is precisely the inverse of the frame time (the frame time represents the period of the rate).
However, in the presence of variable rendering times, a single frame rate cannot accurately represent this variability.
Thus, we choose to focus primarily on frame times in this work.
For those familiar with frame rates, it may be helpful to compute the inverse of a given frame time to gain an intuition for what regular frame rate that frame time corresponds to.
We include frame rates for our constant frame time conditions and summarize our variable frame timing conditions using a $\frac{1}{X}$ convention to provide an intuition for the corresponding frame rates.

More recently (since 2013), variable refresh rate (VRR) technologies, like G-SYNC~\cite{slavenburg2020variable} and VESA Adaptive Sync \cite{callaway2018variable}, have enabled gaming hardware to deliver frames when they are completed.
VRR avoids the potential large magnitude stutter caused by minor frame time variations when VSync is on while still avoiding tearing.
The primary functionality that VRR enables is presenting the variation in frame completion time directly to the display, emitting light to the eyes of the viewer.

Since the introduction and widespread availability of VRR-enabled displays and computing systems, researchers have begun to conduct studies using this paradigm.
For instance, Poth et al.~\cite{poth2018ultrahigh} used G-SYNC to limit the presentation duration of a stimulus to below 10 ms and varied the time in small step sizes to characterize visual processing time.
Others have promoted VRR technology for potential energy savings by reducing power when display updates are not needed~\cite{ko201924,you2020image}.
Another study measured (constant) refresh rates for just noticeable differences (JND) of varying target motion, then leveraged VRR to adjust refresh rate dynamically based on object speed, but did not study the effects of varying refresh rate itself~\cite{Denes2020Motion,mantiuk2021adaptive}.
Jindal et al.~\cite{jindal2021perceptual} added adaptive rate shading to the optimizations for motion, spatial shading rate, and temporal rate but once again neglected to study the effects of varying frame rate in isolation.
This work hints that not all spatio-temporal content has the same presentation time requirements for the equal perception of quality but does not answer whether a single target's motion is the primary driver of the perception of smoothness or stutter in VRR contexts.


Other more relevant works studying VRR technologies' effect on task performance and perception have indicated positive outcomes~\cite{Rihahi2021GSYNC,Watson2019ASYNC}.
While subjects struggled to identify when VRR techniques were enabled, results show significant effects of VRR on both performance and engagement.
Liu et al.~\cite{liu2023effects} studied effects of varied frame timing on quality of experience in 3 real games, as measured by mean observer score.
Instead they relied on a (set of) CPU-based, infinite Fibonacci number counter(s), running in the background, to artificially create uncontrolled variations in frame times.
They find that a 95\% frame rate floor predicts quality of experience across three games.
None of these studies carefully control the sequence, nor the distribution of frame times experienced by the user.
Additionally, they neglected to study any refresh rates above 120 Hz.
We hope our study will fill in some gaps in the current understanding of the effects of VRR on human perception and performance.

\subsection{Motion Artifacts define perceptual smoothness}
\label{sec:motion_artifacts}
To this point, we have alluded to the psychological construct of \emph{perceptual smoothness} and tangentially related it to the motion artifact, \emph{motion blur} in the introduction. 
Prior work has done a fine job of operationally defining perceptual smoothness, concerning object motion, by binning it into various distinct visual percepts: false edges, motion blur, and judder \cite{Denes2020Motion}.
The perception of smoothness comes down to how well people detect (i.e., the visibility of) motion artifacts on displays with different frame rates.
For example, using the visual thresholds of spatio-temporally varying contrast patterns, one can construct a visibility filter in the frequency domain called the \emph{window of visibility} \cite{watson1986window,watson2016pyramid}.
This work was motivated by the fact that the discretization of visual information in the real world adds aliases.
These aliases are characterized by the spatio-temporal sampling rate (e.g., spatial resolution of display pixels or frame rate of displays) and by the modulation pattern of the sampled signals (e.g., pixel fill factor or the hold time of display pixels).
If the window of visibility does not filter out these aliases, the visual system perceives motion artifacts because such aliases are not expected in naturally arising visual inputs. 
The theory of window of visibility has been experimentally measured and verified for displays with constant frame rates \cite{johnson2014motion,hoffman2011temporal,watson2013high} and it provides a comprehensive model of highly controlled laboratory stimuli.

Our work does not attempt to define perceptual smoothness in one way or another (i.e., by the detection of Judder or motion blur) because the presentation of frames vary on a moment-by-moment basis, a different phenomenon relative to presenting frames at constant intervals.
Instead, in the experiments described below, we leave it up to the user to define perceptual smoothness.
Specifically, the user interacts with a gaming environment, and we display the most extreme ends of apparent motion (15 Hz and 360 Hz) to anchor their visual perception for smoothness of motion. 

\subsection{First-person Aiming Performance Measures}
First-person shooters (FPS) are a popular genre of competitively played, high-performance games.
Typically in FPS games, the player uses the mouse to control egocentric view rotation and the keyboard to translate the camera within a scene.
We choose to study the FPS aiming context as it represents an area where high-skill human-computer interaction intersects high-performance rendering and corresponding variable frame timing in an impactful way.
Specifically, we study the camera rotation aspect of FPS aiming in this work, disallowing player translation in the scene to avoid additional confounding factors.
By collecting both perceptual and performance-oriented measures, we investigate how (relatively) small changes in frame timing impact the user experience in competitive FPS gaming.

Previous work has used various metrics for quantifying FPS aiming performance.
Some studies focus on game-specific outcomes tied to the aggregation of many aiming actions, for example, kills, deaths, and assists (as well as their resulting ratios) \cite{shim2011exploratory}.
Others adopt more traditional HCI metrics of time to completion and accuracy (or hit rate) of subjects in completing more atomic tasks \cite{vicencio2014effectiveness}.
Composite score metrics are often used in experiments with more than one primary metric, for example, avoiding being damaged while also damaging targets \cite{claypool2007frame}.
We focus on target elimination time as our metric of interest in this study.

\section{Experimental Design}
We conducted two experiments using systems with similar hardware specifications (see Table \ref{tab:HWspecs}) and a shared, open-source software experimentation platform, First Person Science \cite{Spjut19FPSci, boudaoud2022firstpersonscience} (release v22.05.01), to run and collect data from users.
First Person Science provides a feature whereby an experimenter can specify a target frame time to be used on a per-frame basis. 
We use this virtual sandbox to study the effects of variable frame timing on the user.

In the subsections below, we describe how we generated variable frame time sequences and provide detailed descriptions of the two experiments that focus on the effects of variable frame timing on perception and performance. 
The first experiment is concerned with the visual perception of smoothness of motion for sphere-shaped targets in a first-person targeting task. 
The second experiment builds on the first by selecting subsets of equally smooth (as measured by experiment one) frame time sequences and comparing performance across those groupings. 
Lastly, we include a description of the participants in each user study. 

Based on our understanding of the impacts of frame time and its variation from prior art and pilot studies in FPS aiming, we expect that:

\begin{enumerate}
    \item Players can perceive 4-12 ms variations in frame time, across a range of base frame rates (30-240 Hz)
    \item Players incur additional performance impacts in FPS targeting under perceptually equivalent VFT conditions
\end{enumerate}

We base these hypotheses on prior studies indicating that changes in frame time are highly perceptible.
Furthermore, we suspect that VFT will have additional performance impacts as prior art had found VRR techniques such as G-SYNC improved player performance even when the change was not perceived \cite{Watson2019ASYNC}.

\begin{table}
    \centering
    \begin{tabular}{l|l}
         \textbf{Parameter} &  \textbf{Value}\\
         \hline
         Display &  360 Hz Alienware AW2521H\\
         Resolution & 1920 $\times$ 1080\\
         G-SYNC (VRR) & Enabled\\
         VSync & Disabled \\
         Viewing Distance & \(\sim \)70 cm\\
         \hline
         CPU & Intel i7-9700k @ 3.60 GHz \\
         GPU & NVIDIA RTX 2080Ti \\
         RAM & 32GB DDR4\\
         \hline
         Mouse & Logitech G Pro Wireless\\
         Mouse DPI & 800 \\
         Keyboard & Logitech G Pro YU0039\\
         Click-to-Photon Latency & 20 ms @ 120 Hz (Sec. \ref{sec:Latency})\\
    \end{tabular}
    \caption{Hardware specifications for the systems used in our experiments.}
    \label{tab:HWspecs}
\end{table}

\subsection{Variable Frame Time Approach}
\label{sec:VariableFrameTime}
Across our experimental conditions, subjects experienced a range of constant and variable frame time sequences over which they ranked smoothness and attempted to complete targeting tasks as quickly as possible.
In deciding which type of variable frame timing (VFT) scheme to use, we had many different candidates.
Initially, we considered a "1:N stutter" scheme where a series of N frames occur at constant frame timing, followed by a single (or small set of) frames(s) that is (are) longer/shorter.
This model creates a rapid change in frame timing that emulates a single performance hitch or unpredictable rendering which are common in real games.
However, this approach is not desirable for the purpose of this study, in particular for experiment 2, because participants may accomplish the task in a time window adjacent to where the stutter occurs and thus the VFT scheme would not impact performance.

\begin{figure}
    \centering
    \includegraphics[width=\columnwidth]{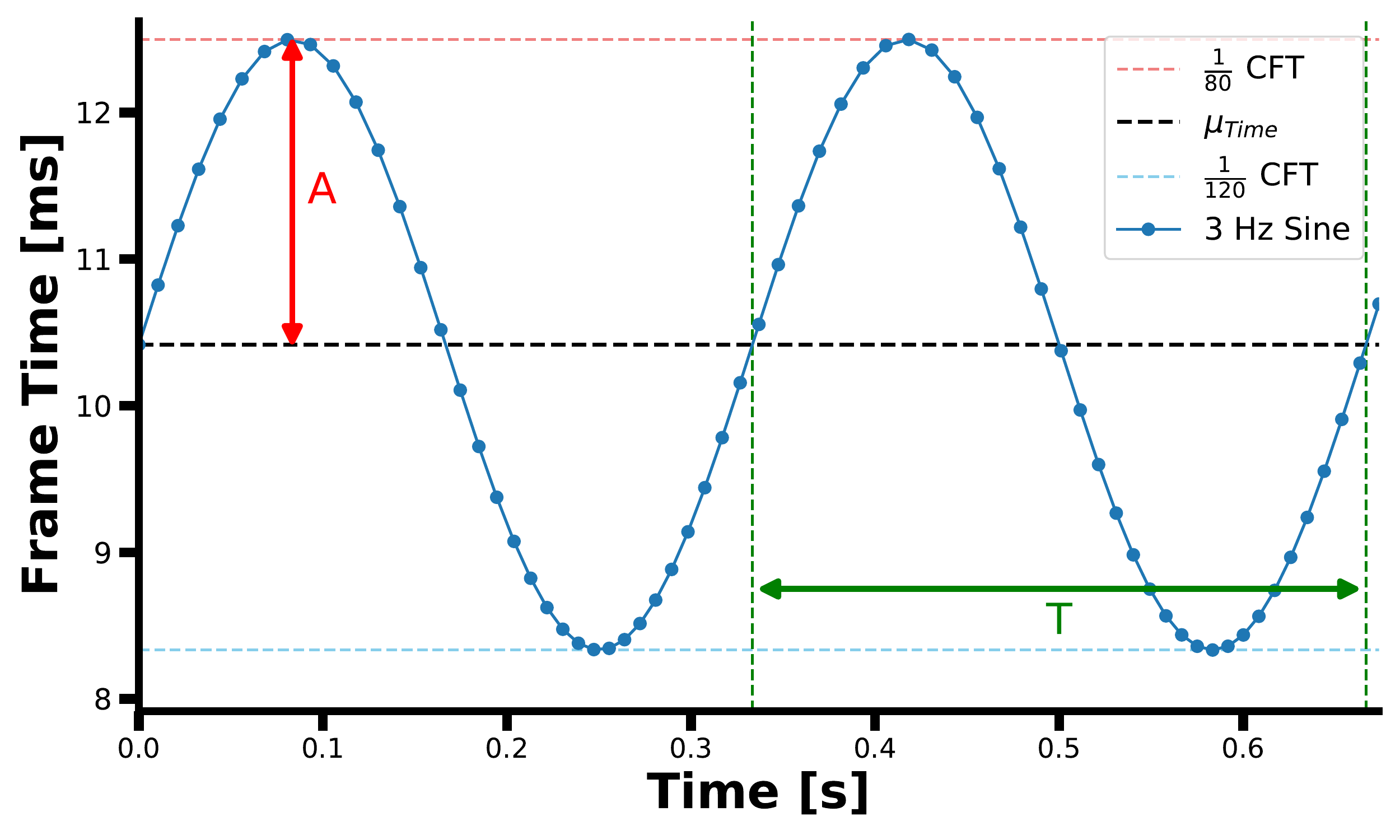}
    \caption{Two periods of an example 3 Hz sine wave of varying frame time over time with 2 ms amplitude (A), 333.33 ms period (T), and time-weighted mean frame time ($\mu_{\textrm{Time}}$ = 10.42 ms). Note that the scatter points (and their relative order over time) overlaid on the waveform are the frame times users experience when participating in the task. The dashed red/blue lines indicate the minimum and maximum frame duration denoted $(\frac{1}{120},\frac{1}{80})$ (see the supplement for more details). 
    }
    \label{fig:VariableFrameTime}
\end{figure}

We elected to use a sine wave sampling scheme of frame time over time to produce our VFT sequences.
The signal had a frequency of 3 Hz, and the waveform spanned between pairs of CFTs also tested.
We chose a sine wave of this frequency because it gives a smoothly varying VFT over time.
3 Hz is slow enough that the waveform will have enough samples with frame times as high as 33.3 ms and fast enough that a player will experience at least one full waveform within an expected task completion time of over 500 ms\footnote{A pilot study considering a few different frequencies showed minor differences in perception.Thus we selected only a single frequency.}.
Fig.~\ref{fig:VariableFrameTime} plots two periods of an example waveform and shows an example of how a thoughtful sampling of the intended signal is required to avoid pitfalls.
The supplemental video visualizes this approach using lower frame times (<30 Hz) and a lower VFT sine wave period (1 Hz), to make the effects more clearly visible on common displays.

\subsection{Experiment 1: Perceptual Smoothness}
The high-level flow of experiment 1 is summarized in the left-most portion of Fig. \ref{fig:expflow}.
Participants were instructed to rate the smoothness of the gameplay on a scale of 1-7 as they performed a secondary first-person targeting task.
This task required participants to destroy as many moving targets on screen as possible in the allotted time. 
To destroy a target, a participant aligns a reticle (crosshair) over a moving target and clicks once. 
Following a valid mouse click, the target is destroyed and automatically respawns at a random location in the scene. 
Four targets were presented simultaneously and moved in random directions across the screen. 
Participants could not translate (move) their camera within the 3D scene but could rotate the view direction by moving the mouse, reproducing aiming controls typical of FPS gaming.

\subsubsection{Trial flow}

\begin{figure*}
    \centering
    \includegraphics[width=0.49\textwidth]{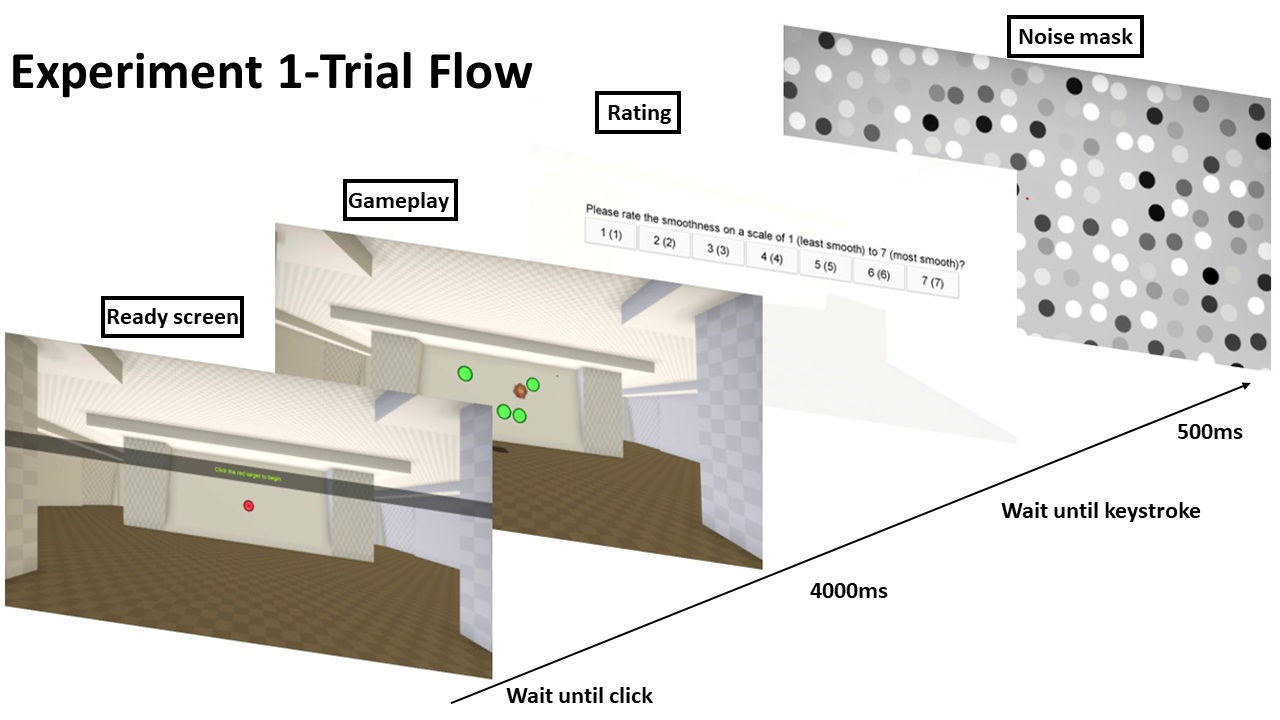}
    \includegraphics[width=0.49\textwidth]{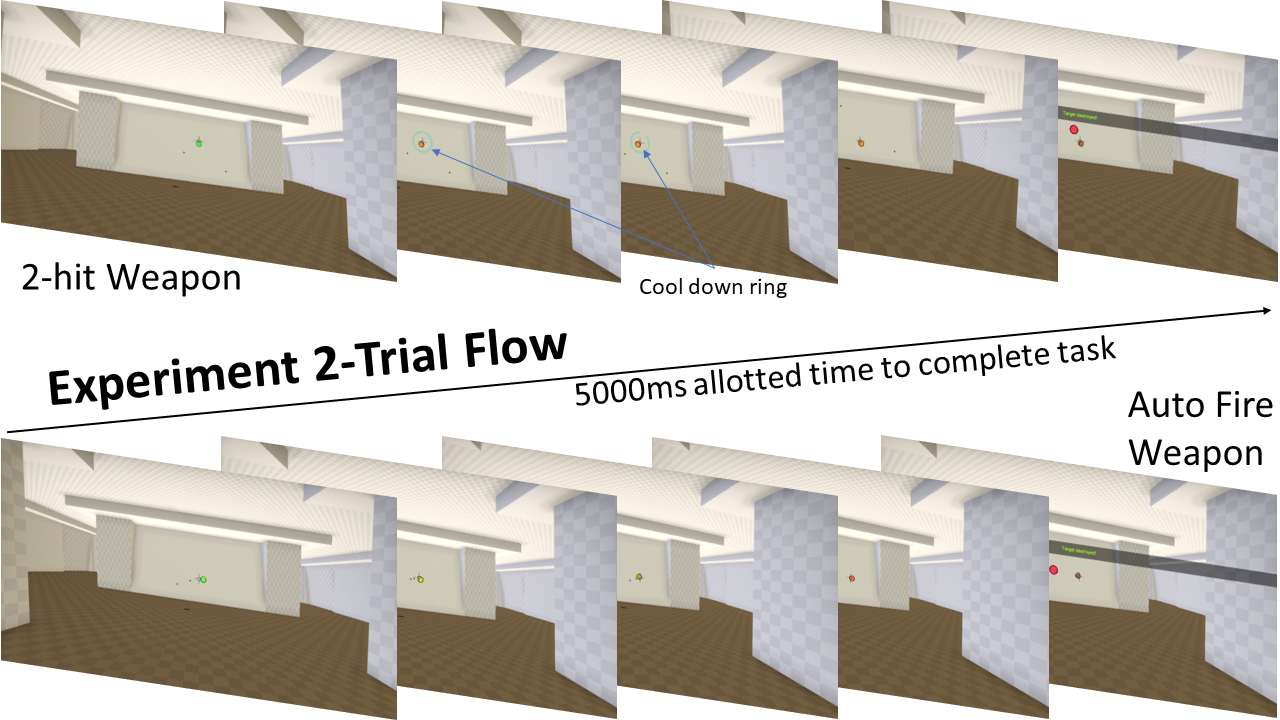}
    \caption{Trial flow for experiment 1 (left) and experiment 2 (right). In experiment 1, only one weapon type results in a single flow. Note that the visual noise mask is presented in the final stage of the experiment 1 flow. In experiment 2, participants utilized two weapon types, and the damage sequence is depicted across frames of two trials.}
    \label{fig:expflow}
\end{figure*}

Each trial began with a red reference target centered on the screen, with the reticle’s default position set to the center of this target. 
A message was overlaid on-screen stating: “Click the red target to begin.” 
When participants were ready to begin a trial, they clicked on the reference target. 
Afterward, four green targets appeared on the screen. 
The targets moved randomly within a predefined region in world-space coordinates for four seconds. 
We instructed participants to try their best to destroy the targets in each trial and warned them that their data would be discarded if the number of targets destroyed did not meet an unspecified threshold. 
We kept this threshold vague to ensure that participants focused on the judgment of smoothness while simultaneously doing the secondary task of destroying the targets. 
We decided to have the targets respawn infinitely to ensure that participants experienced the complete sequence of frame times before making a perceptual judgment.

After four seconds of gameplay elapsed, a rating screen appeared. 
Participants indicated their perception of smoothness on a Likert scale ranging from 1-7, with 1 mapping to the least smooth perceptual experience and 7 mapping to the smoothest perceptual experience. 
Participants marked their responses by pressing the appropriate number key on the keyboard. 
To end the trial, a dynamic visual noise mask (run as a 2D shader) \cite{ap2015sorbet} would appear on the screen for half a second. We added the mask in an attempt to remove any carryover effects or motion adaption from the previous trial. 
This flow is summarized visually on the left side of Fig. \ref{fig:expflow}.

Before the experiment started, participants were allowed to adjust their mouse sensitivity to their desired setting. 
The mouse sensitivity adjustment task was similar in design to the description above, except it ended once all 4 targets were destroyed with no time limit. 
Afterward, participants were provided task instructions and practice trials to get comfortable with the experiment design.
The instructions consisted of two separate trials.
The first trial lasted ten seconds instead of four, with the game rendered at 15 Hz.
Participants were instructed to pay attention to the smoothness of the gameplay, as this visual experience represented a smoothness rating of 1.
Next, they experienced a trial at 360 Hz for ten seconds and were told that this visual experience mapped to a smoothness rating of 7.
As a note, we did not include 15 Hz or 360 Hz frame rates in the actual experiment. 
After being introduced to the rating scale, participants experienced five warm-up trials lasting four seconds each. 
The warm-up trials consisted of variable frame time sequences and constant frame time sequences and were designed to introduce the user to the primary task of the experiment. 

\subsubsection{Stimuli}
We utilized four instances of a single type of target in each trial. 
The target was green and had a world-space diameter of 0.5 m.
Note that we did not calculate degrees of visual angle subtended by the target on the retina as we did not use a chin rest to control for viewing distance.
Additionally, the apparent size of the target changed slightly on-screen as it moved due to perspective projection.
The target moved at a constant world-space speed of 3 m/sec.
At some point, the target changed direction every 700-1000 ms (uniformly sampled), with the new direction also sampled uniformly from all possible directions.
The targets spawned in a 0.5 x 2 x 2 m box centered 7.5 m away from the player spawn position.
Those targets moved within a larger 1 x 2.3 x 6 m box centered on the same position as the spawn bounding box.

We utilized a red circle for the reticle. 
Participants could adjust their mouse sensitivity to their desired setting (in $^\circ$/mm), and the mouse DPI was set to 800. 

\subsubsection{Conditions}
Experiment 1 employed four conditions, one constant frame time and three variable frame time conditions that differed in amplitude (see Section \ref{sec:VariableFrameTime} above).
The constant frame time condition was comprised of eight levels.
From the longest frame duration to the shortest, we included: 1/30, 1/34, 1/40, 1/48, 1/60, 1/80, 1/120, and 1/240 s.
We selected these frame duration levels based on two factors. 
First, we wanted to adequately sample the domain of monitor refresh rates available in the wild. 
Older displays and TV content update at 30 Hz refresh rates, while new monitors can run up to 360 Hz, but we only tested up to 240 Hz as this is what is widely available in the gaming market.
Second, the VFT conditions, using a sinusoidal variation, were composed of frame sequences that oscillated periodically between two frame times; for example, between 1/240 s and 1/120 s.
Because the user is experiencing (at most) a frame time equivalent to a 120 Hz frame rate, we wanted to include that constant frame rate in our experiment design.  

The three VFT conditions sampled the same domain of frame times as the constant frame time condition but differed from one another in the amplitude (A) of the sine wave used to generate the sequence of frame times.
The amplitude and the min and max frame duration from both the constant and variable frame time conditions are provided in Table \ref{tab:conditions}.


\begin{table*}[]
    \centering
    \begin{tabular}{l|l|l}
        \textbf{Amplitude} & \textbf{Count} & \textbf{Frame Duration (min, max) [s]} \\
        \hline
        0 ms (constant) & 8 & $\frac{1}{240}$, $\frac{1}{120}$, $\frac{1}{80}$, $\frac{1}{60}$, $\frac{1}{48}$, $\frac{1}{40}$, $\frac{1}{34}$, $\frac{1}{30}$\\
        2 ms & 7 & ($\frac{1}{240}$,$\frac{1}{120}$), ($\frac{1}{120}$, $\frac{1}{80}$), ($\frac{1}{80}$, $\frac{1}{60}$), ($\frac{1}{60}$, $\frac{1}{48}$), ($\frac{1}{48}$, $\frac{1}{40}$), ($\frac{1}{40}$, $\frac{1}{34}$), ($\frac{1}{34}$, $\frac{1}{30}$)\\
        4 ms & 6 & ($\frac{1}{240}$, $\frac{1}{80}$), ($\frac{1}{120}$, $\frac{1}{60}$), ($\frac{1}{80}$, $\frac{1}{48}$), ($\frac{1}{60}$, $\frac{1}{40}$), ($\frac{1}{47}$, $\frac{1}{34}$), ($\frac{1}{40}$, $\frac{1}{30}$)\\
        6 ms & 5 & ($\frac{1}{240}$, $\frac{1}{60}$), ($\frac{1}{120}$, $\frac{1}{48}$), ($\frac{1}{80}$, $\frac{1}{40}$), ($\frac{1}{60}$, $\frac{1}{34}$), ($\frac{1}{48}$, $\frac{1}{30}$)\\
    \end{tabular}
    \caption{Experiment 1 conditions are ordered by the amplitude of frame time variation. Constant (0 ms) conditions, with equal minimum and maximum frame duration, are represented by a single number instead of a pair. All frame duration is presented as fractions in seconds. The count column represents the number of levels tested in each of the four conditions.}
    \label{tab:conditions}
\end{table*}

In total, there were 26 levels across the four conditions. Participants encountered each of the 26 levels 11 times totaling 286 trials. To account for order or carryover effects, we presented the 26 levels in random order within a single block (of 26 trials). Participants, therefore, encountered 11 blocks of 26 trials so that no block contained two trials of the same level. This experiment took approximately 1 hour to complete. 

\subsubsection{Participants}
A total of 10 participants completed experiment 1. However, one participant's data were excluded from analysis because their responses were overly noisy (i.e., the standard error of the mean for their response at each level was 2-3$\times$ as large as the other participants). All participants signed an informed consent form approved by the IRB (Protocol \#00059738). 


\subsection{Experiment 2: Targeting Performance}
Experiment 2 was a performance-based task focused on the effect of variable frame timing while controlling for perceptual smoothness (based on experiment 1 results) on destroying a single target with two weapon configurations: 2-hit and Autofire.
The first weapon configuration was a 2-hit weapon that required participants to click twice on the target to destroy it.
The second weapon configuration was set to Autofire, which required a single mouse click followed by a hold to fire repeatedly (15 shots/s).

Our selection of constant and variable frame time sequences was based on the perceptual results from experiment 1.
Specifically, we paired VFT sequences and constant frame time sequences that were perceived as equally smooth in experiment 1.
With this framework in mind, we can assess whether the VFT sequences impact targeting performance relative to a constant frame time sequence despite being perceived as equally smooth.  

\subsubsection{Trial flow}
A trial in experiment 2 was similar in structure to a trial in experiment 1 but with a few important differences.
First, a single green target appeared on screen as opposed to four.
Second, the participants were instructed to destroy the target as quickly as possible within five seconds to complete the task.
The trial would end if either five seconds passed or the target was destroyed, whichever came first.
At the end of the trial, a message would appear stating either, “Target destroyed!” or “Target not destroyed.”
The subsequent trial would begin with the following message, “Click the red target to begin.”

\subsubsection{Stimuli}
\begin{figure}
    \centering
    \includegraphics[width=0.82\columnwidth]{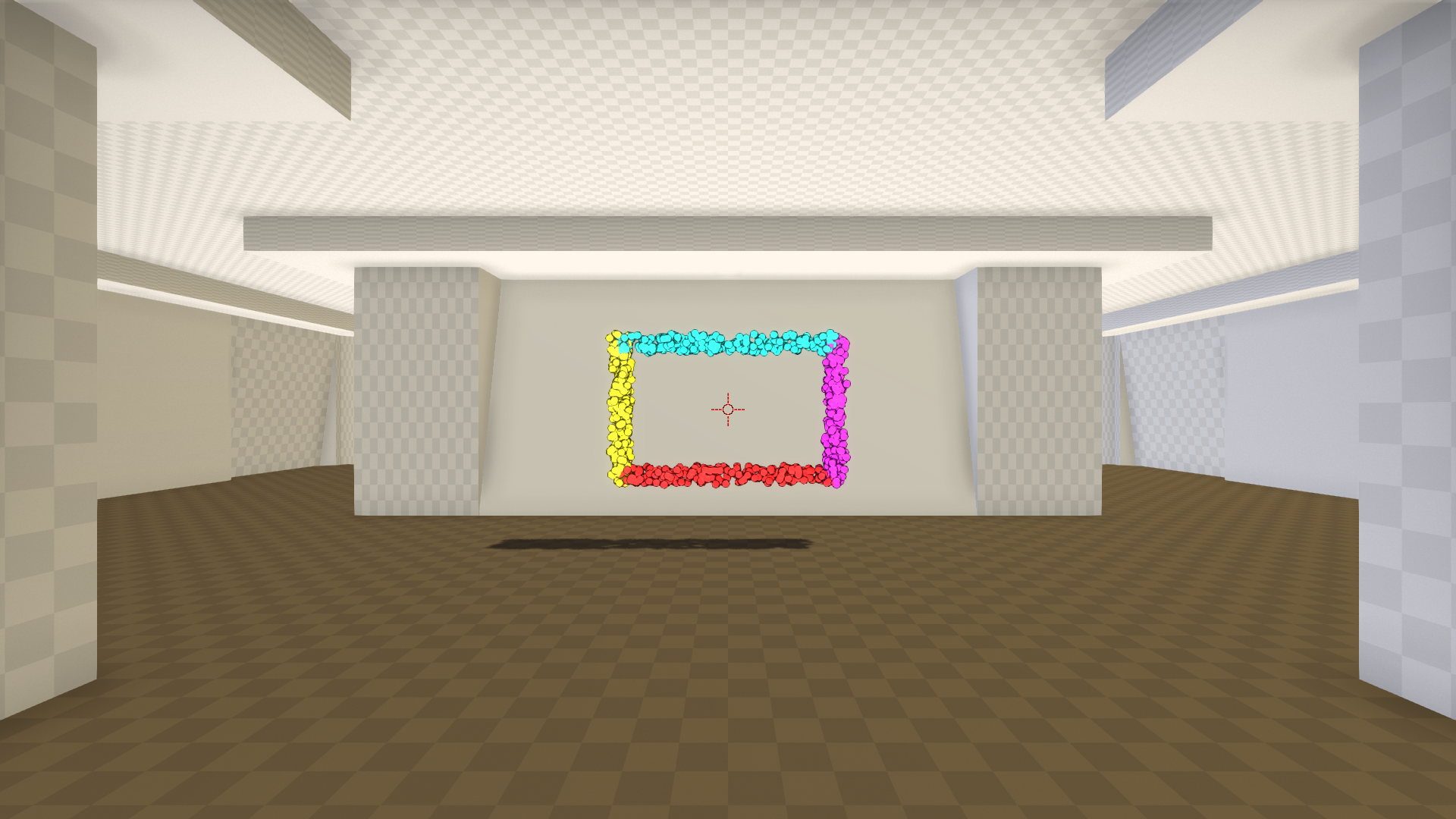}
    \includegraphics[width=0.17\columnwidth]{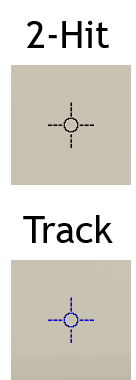}
    \caption{(Left) Visualization of the experiment 2 spawn regions centered on the initial player aim direction. The top (cyan) and bottom (red) regions were 0.5 x 0.2 x 2.6 m and the left (yellow) and right (magenta) regions were 0.5 x 1.8 x 0.2 m. An equal number of targets spawned in each region within each condition. The reticle was drawn in red at the center of the screen. (Right) Different colored reticles used for the 2-hit (black) and track (blue) weapon configurations.}
    \label{fig:exp2spawn}
\end{figure}

Participants were instructed to track and destroy a single green target.
The target was 0.25 m in diameter and moved at a constant speed of 2 m/s. 
The target utilized the same direction change algorithm as experiment 1 with an equal motion change time range of 700-1000 ms. 
Given that this task was performance-based, we wanted to avoid the worst possible edge case (and similar cases) where the target spawns, by chance, directly over the player reticle, as this would unduly influence task performance.
Therefore, we created four different rectangular spawn regions that comprise the perimeter of the bounding box of target motion.
These regions are visualized in Fig. \ref{fig:exp2spawn}.
The top and bottom regions were 0.5 x 0.2 x 2.6 m and the side regions were 0.5 x 1.8 x 0.2 m.
The target movement bounds were the same as in experiment 1 (a 1 x 2.3 x 6 m box centered 7.5 m away from the user and completely containing the spawn bounds). 

When the 2-hit weapon configuration was selected, the reticle was drawn in black, and when the Autofire weapon was used, the reticle was drawn in blue, as shown on the right side of Fig. \ref{fig:exp2spawn}, with the intent to inform the player which weapon was active.
The 2-hit weapon fired at 4 Hz and required two shots on target to destroy. 
Each shot was followed by a cooldown period, indicated by a ring that appeared on the screen surrounding the reticle.
It changed color from blue to yellow during the cooldown, and once it disappeared, participants could fire the weapon again, as shown in Fig. \ref{fig:expflow} right. 
The Autofire weapon fired 15 shots/s and required 10 hits to destroy the target. 
Neither weapon supported a reload action, and both had infinite ammunition. 
Lastly, participants could adjust their mouse sensitivity to their liking (in $^\circ$/mm), and the mouse DPI was fixed to 800.

\subsubsection{Experiment conditions}
In this experiment, we employed three constant frame time levels and five variable frame time levels from experiment 1.
Based on the perceptual results from experiment 1 (Sec.~\ref{sec:exp1}), we grouped constant frame time sequences with VFT sequences that were, on average, perceived as equally smooth.
The first pairing contained a constant frame time sequence at 1/120 s and a VFT sequence that oscillated between 1/240 and 1/80 s (A = 4 ms). 
The second grouping contained the constant frame duration of 1/60 seconds and three VFT sequences: 1) 1/120 to 1/48 (A = 6 ms), 1/80 to 1/48 (A = 4 ms), and 1/60 to 1/48 (A = 2 ms).
The last pairing considered the constant frame time sequence of 34 Hz and the VFT sequence oscillating between 1/40 to 1/30 seconds (A = 4ms). 
Together, these groupings constitute 3 conditions for experiment 2.
For each of the variable and constant frame time sequences in the 3 conditions, participants completed the task with the 2-hit and Autofire weapon configurations.
There were 16 unique frame time-by-weapon configuration combinations across the three conditions.

We utilized a blocked design where participants completed all trials for one frame time sequence by weapon configuration combination (e.g., 1/60 s constant frame time sequence with the 2-hit weapon) before moving on to the next block of trials.
Within a given block, participants completed 5 mini-blocks. 
Each mini-block contained eight trials, 2 trials for each of the four target spawn regions (left, right, top and bottom). 
We discarded the first mini-block as training from the initial 40 trials. 
We analyze 32 trials per block totalling 640 trials per subject in just over an hour. 

To account for order effects, we controlled the sequence of blocks participants encountered throughout the experiment.
First, we took all permutations of the three conditions (6) and randomly shuffled the ordering of the constant and variable frame time blocks within each of the conditions.
Then for each permutation, we created a sequence of blocks where the odd-numbered blocks (i.e., blocks 1, 3, …, 15) used the 2-hit weapon, and the even-numbered blocks used the Autofire weapon. 
We also include a separate order where participants utilized the Autofire weapon on all the odd-numbered blocks and the 2-hit on the even-numbered blocks. 
Together, this amounted to 12 orderings of blocks. 

\subsubsection{Participants}
A total of 15 participants completed experiment 2. All participants signed an informed consent form that was approved by the IRB (Protocol \#00059738). 

\begin{figure*}
    \centering
    \includegraphics[trim=160 0 160 0, clip,width=\textwidth]{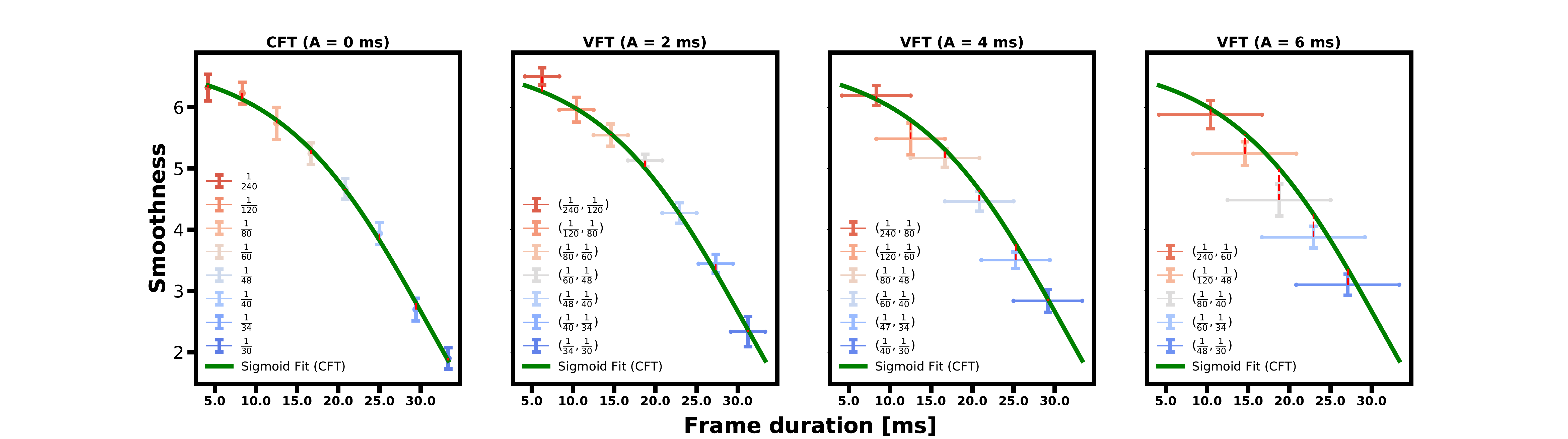}
    \caption{Perception of smoothness plotted as a function of time-weighted mean frame duration for each of the four conditions tested in experiment 1. Participants utilized a rating scale from 1 (low) to 7 (high) to indicate their perception of smoothness during gameplay. (Left) Mean smoothness ratings across subjects for each of the eight frame duration levels tested in the constant frame time condition. Error bars represent the standard error of the mean. A sigmoid function (see Equation \ref{eq:sigmoid}) is fit to the pooled participant smoothness scores for constant frame rates and is superimposed over all plots in the figure (green lines). We computed MSE as a measure of goodness of fit, and red dotted lines depict prediction errors of the fit with respect to observed data. (From middle left to right) 2/4/6 ms amplitude VFT condition: smoothness scores for each of the levels presented to participants are plotted against the time-weighted mean frame duration of each VFT sequence (i.e., the x-coordinate where vertical error bars intersect the horizontal lines).}
    \label{fig:exp1_fit}
\end{figure*}

\section{Results and Analysis}
\subsection{Experiment 1-Perception}
\label{sec:exp1}
When comparing the frame time sequences (levels) within the same condition, perceived smoothness monotonically decreased as frame duration increased.
This relationship held across all four conditions tested in experiment 1 and is shown in Fig. \ref{fig:exp1_fit} where perceived smoothness is plotted against time-weighted mean frame duration (the x-coordinate where vertical error bars project out in the right three panels). 
This shows that the findings from prior work \cite{Denes2020Motion} continue to hold even when frame time varies.
The smoothness rating changed almost linearly as the time-weighted mean frame duration increased, with the slope being slightly less steep at the short frame times.
It seems that meaningfully different smoothness existed across the tested range of frame times.
Additionally, our reference stimuli effectively anchored the perceived smoothness at 15 and 360 Hz to the smoothness ratings of 1 and 7, respectively.

To examine the effect of frame time variation, we fit a sigmoid function to the observed data in the CFT condition (leftmost panel in Fig. \ref{fig:exp1_fit}) using a least-square criterion,
 \begin{equation}
     f_{\alpha,x_0, \beta, C}(x) = \frac{\beta}{1+e^{-\alpha(x-x_0)}}+C
     \label{eq:sigmoid}
     \vspace{4mm}
 \end{equation}
yielding the following fit parameters: $\alpha=-116.983$, $\beta=7.63$, $x_0=0.0306$, $C=-1.091$ with a mean squared error (MSE) of  0.0458.

We then overlaid the fitted sigmoid curve on the VFT sequences comprising the other three conditions, as depicted by the green curve re-plotted over the observed data in each of the right three panels of Fig. \ref{fig:exp1_fit}.
For each VFT sequence (horizontal lines in Fig. \ref{fig:exp1_fit}), one can visually search for the CFT yielding equivalent predicted smoothness by finding the point on the green curve with an equal smoothness rating\footnote{The green curve interested all VFT sequences except for the (1/240-1/120s) level. 
We exclude this data point from analysis but revisit it in the discussion.}.
In other words, we identify where the green curve intersects the horizontal line.
This equivalently smooth CFT tends to be closer to the longest frame duration in the given VFT sequence (the right edge of the horizontal line) than the shortest frame duration in the sequence (the left edge of the same line).

\begin{figure*}
    \centering
    \includegraphics[width=0.7\textwidth]{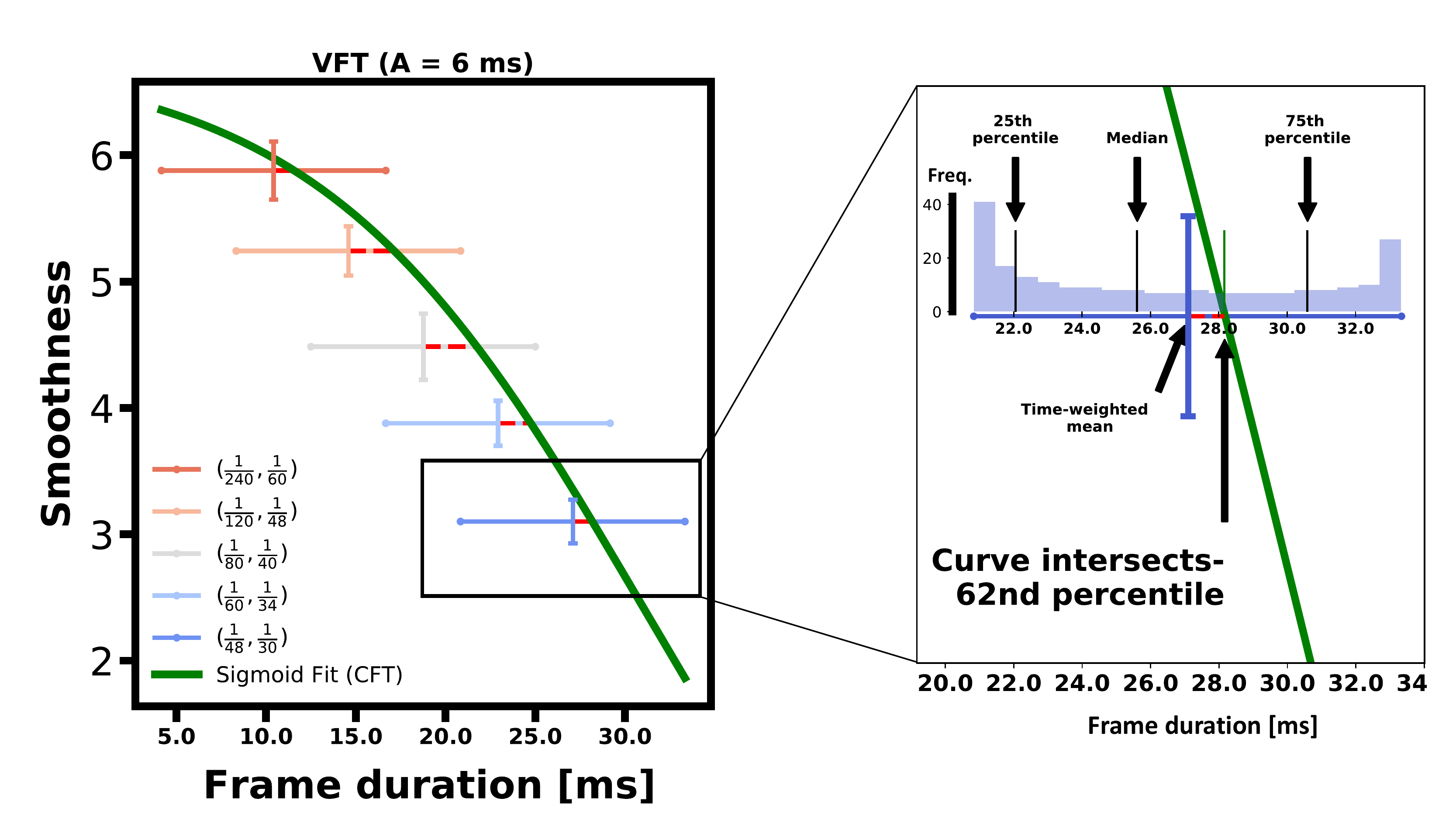}
    \caption{Perception of smoothness in the 6 ms VFT condition (A = 6 ms). (Left) The sigmoid fit to the constant frame duration data (green curve) is overlaid on top of horizontal bars representing the range of frame times for each of the five levels in the A = 6 ms VFT condition. Horizontal dashed-red lines extend from the time-weighted mean frame duration of each VFT sequence to the intersection point of the green curve and the horizontal lines. We consider these intersection points as equivalently smooth CFTs for corresponding VFT sequences. The inset figure to the right depicts the frequency distribution of frame times in the VFT sequence oscillating between (1/48, 1/30) frame times. Vertical bars overlaid on the distribution highlight the 25th, median (50th), and 75th percentiles and the time-weighted mean frame duration. Additionally, the last vertical bar highlights the percentile rank of the frequency distribution corresponding to the equivalently smooth CFT.}
    \label{fig:exp1_percentile}
\end{figure*}

In applying the same logic from the above analysis to each of the three VFT conditions, we found that the amplitude of frame time variation had a systematic effect on the equivalently smooth CFT (Fig. \ref{fig:exp1_percentile}).
When the amplitude of frame time variation was small (2 ms), the equivalently smooth CFT was not strongly biased toward the longer frame time in the VFT condition.
Additionally, it tended to cluster around the time-weighted mean frame duration of each VFT sequence. 
When the amplitude of frame time variation was large (6 ms), the equivalently smooth CFT became strongly biased toward the longer frame time in the VFT sequence (Fig. \ref{fig:exp1_percentile} left).
We can conceptualize this bias another way by considering the equivalently smooth CFT as a percentile for the distribution of frame times in the VFT sequence under consideration, as exemplified in the inset plot of Fig. \ref{fig:exp1_percentile} for the sequence (1/48, 1/30).

\begin{figure}
    \centering
    \includegraphics[width=\columnwidth]{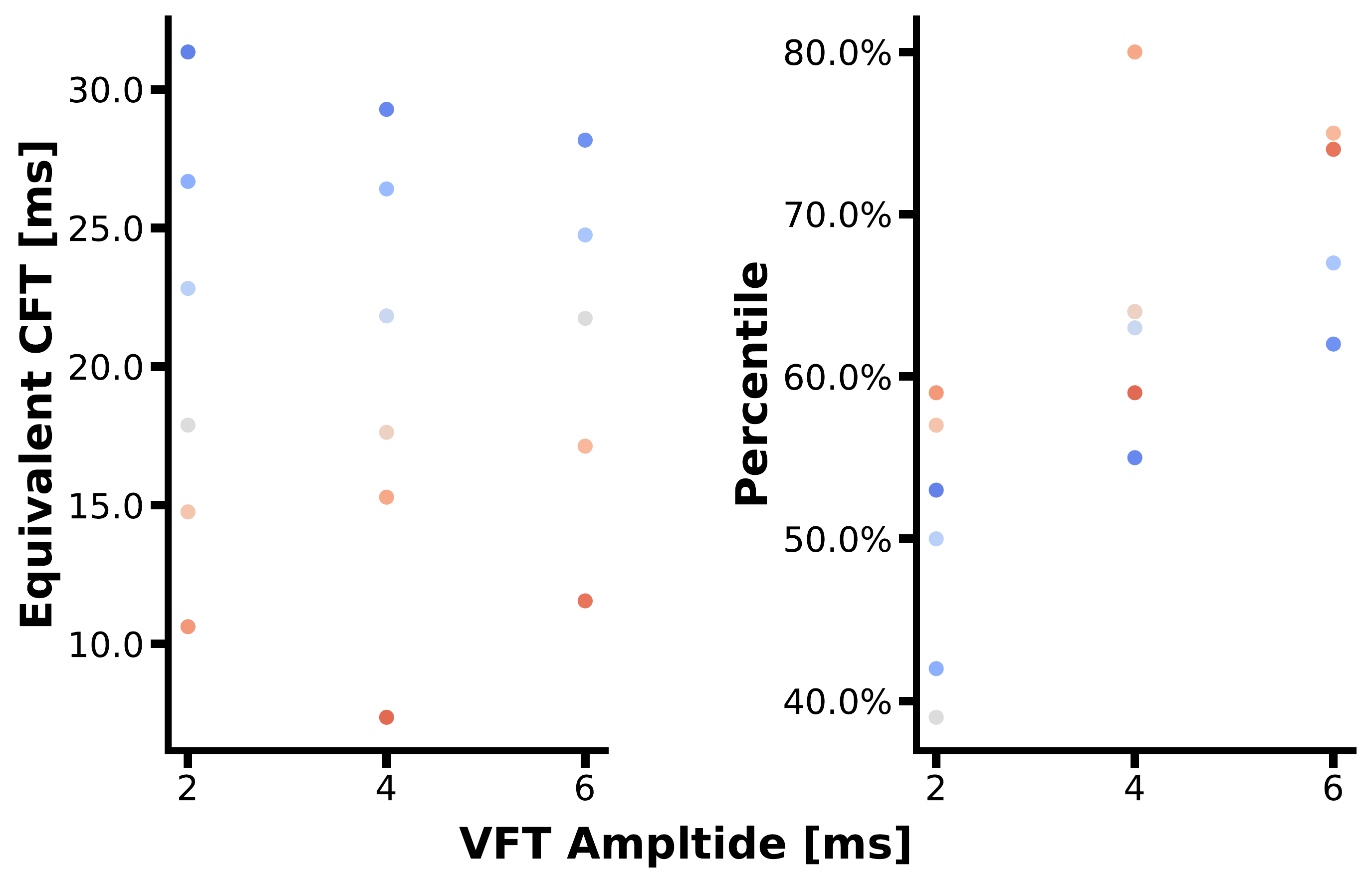}
    \caption{(Left) The scatter plot demonstrates a poor correlation between VFT amplitude and the smooth CFT metric. (Right) Improved positive correlation of VFT amplitude when equivalently smooth CFT is converted to a percentile rank for the distribution of frames in its respective VFT sequence. Note that the colors of scatter points map directly to the colors in the legend of \ref{fig:exp1_fit}.}
    \label{fig:exp1_percentiles_vs_frame_duration}
\end{figure}

After converting the equivalently smooth CFT to a percentile rank for each VFT sequence in a particular condition, we see that there is a lower variance around a higher mean percentile rank in the 6 ms condition relative to the 4 ms and 2ms conditions (Fig. \ref{fig:exp1_percentiles_vs_frame_duration}, right).
However, when we consider the equivalently smooth CFTs, this relationship across VFT condition disappears as shown in Fig. \ref{fig:exp1_percentiles_vs_frame_duration}, left.
The analysis demonstrates that the perceived smoothness in VFT becomes dominated by the visual stimulation from longer frames, and the bias is stronger and more consistent when the amplitude of the variation is wider.

\begin{table*}
    \centering
    \begin{tabular}{c|c|c|c}
         \textbf{Condition} & \textbf{MSE Weighted Mean} & \textbf{MSE Percentile} & \textbf{Reduction in MSE} \\
         \hline
         2 ms & 0.0351 & 0.0334 & 5\% \\
         4 ms & 0.0532 & 0.0537 & -0.95\% \\
         6 ms & 0.0053 & 0.0043 & 19.7\% \\
    \end{tabular}
    \caption{Comparison of sigmoid MSE for time-weighted mean and percentile-based fits. Note that comparisons should occur within conditions, not across them, as the number of points fit by the 4-parameter sigmoid (see Eq. \ref{eq:sigmoid}) was different between conditions.}
    \label{tab:percentileMSE}
\end{table*}

To confirm this, we fit two sigmoids for each amplitude condition. 
The first sigmoid was fit on the time-weighted mean frame duration of each VFT sequence, and the second one was fit on the mean percentile of the scatter points in Fig. \ref{fig:exp1_percentiles_vs_frame_duration} (right). 
Overall, we observe more accurate fits when using the mean percentile instead of the time-weighed mean of the frame time series (Table \ref{tab:percentileMSE}).
This indicates that, particularly for larger frame time variations, using a model based on a fixed (high) percentile of frame times better predicts perceived smoothness than time-weighted mean frame time.

\subsection{Experiment 2-Performance}
\subsubsection{Figures of merit and Analysis}
We looked at time to completion to evaluate the effect of variable frame timing on performance. 
As a baseline check, we also looked at the average number of trials with the target destroyed per VFT or CFT sequence. 
The task, by design, was easy to accomplish, so the proportion of trials in which the target was successfully destroyed served as a measure to rule out outliers.
We chose a threshold of 75\% of trials with the target destroyed to rule out outliers.
Two participants did not meet this threshold and were excluded from the statistical analysis, leaving 15 participants.
Our primary focus was to assess the time to complete the task. 
We reasoned that if VFT were to affect performance, then this would be reflected as slower completion times for trials with varying frame times relative to trials with constant frame times. 


For time to complete the task, we coupled two statistical analyses to assess different but related hypotheses.
First, we conducted a three-way repeated measure ANOVA (\emph{2 weapon configurations} $\times$ \emph{3 smoothness levels} $\times$ \emph{2 frame timing regimes}) to determine if there was an interaction between perceptual smoothness, weapon configuration, and VFT on time to destroy the target.
We employed a fully crossed design where our focal, or moderating, variable was weapon configuration.
We hypothesized that the effect of weapon type on time to completion would depend on both the smoothness of the gaming experience and whether the frame duration was varying or constant. 
The smoothness variable was comprised of three levels based on the findings from experiment 1: low smoothness (1/34 s, and (1/40 s-1/30 s)), average smoothness (1/60 s, and (1/80 s-1/48 s)), and high smoothness (1/120 s, and (1/240 s-1/80 s)). 
Note that each variable frame time sequence has an amplitude of 4 ms, so the only factors changing are the sequence's minimum and maximum frame duration, which are directly related to the smoothness experience.
The variable frame time factor has two levels: constant frame time (1/34 s, 1/60 s, and 1/120 s) and variable frame time ((1/40 s-1/30 s), (1/80 s-1/48 s), and (1/240 s-1/80 s)). 
The last factor, weapon configuration, has two levels: 2-hit and Autofire.

Our second analysis focused on how the sine wave's amplitude influenced performance.
We conditioned our analysis on a single smoothness score and investigated the interaction between weapon configuration and amplitude of the sine wave.
The first factor, amplitude of the sine wave, consisted of 4 levels: 1/60 s (A = 0 ms), 1/60 s-1/48 s (A = 2 ms), 1/80 s-1/48 s (A = 4 ms), 1/120 s-1/48 s (A = 6 ms).
The second factor, weapon configuration, maintained the same two levels as the previous analysis.
In short, we ran a fully crossed two-way repeated measure ANOVA (\emph{2 weapon configurations} $\times$ \emph{4 amplitudes}).

Statistically significant main effects in both analyses were followed up with multiple paired-sample t-tests with Bonferroni adjustments made at an alpha level of 0.05.
Specifically, the p-values reported in the experiment 2 results section are adjusted using a Bonferroni correction. 
We also wrangled the data for each participant to conform to the assumptions of the ANOVA test.
First, we filtered out the data from each block's first mini-block (8 trials) to focus on steady-state performance and exclude outliers that may have resulted from adapting to the task.
Afterward, we computed the mean time to complete the task with the remaining 32 trials (although some means were based on smaller sample sizes if participants did not complete the task on a particular trial).
We then applied a natural log transformation (base $e$)  to the mean value obtained for each observer to ensure that the residuals were normally distributed.
Note that we report the descriptive statistics (in both our figures and text) in the original units despite running the analyses on the transformed data for ease of interpretation (i.e., equal length standard error whiskers around means on bar charts).

\begin{figure*}
    \centering
    \includegraphics[height=52mm]{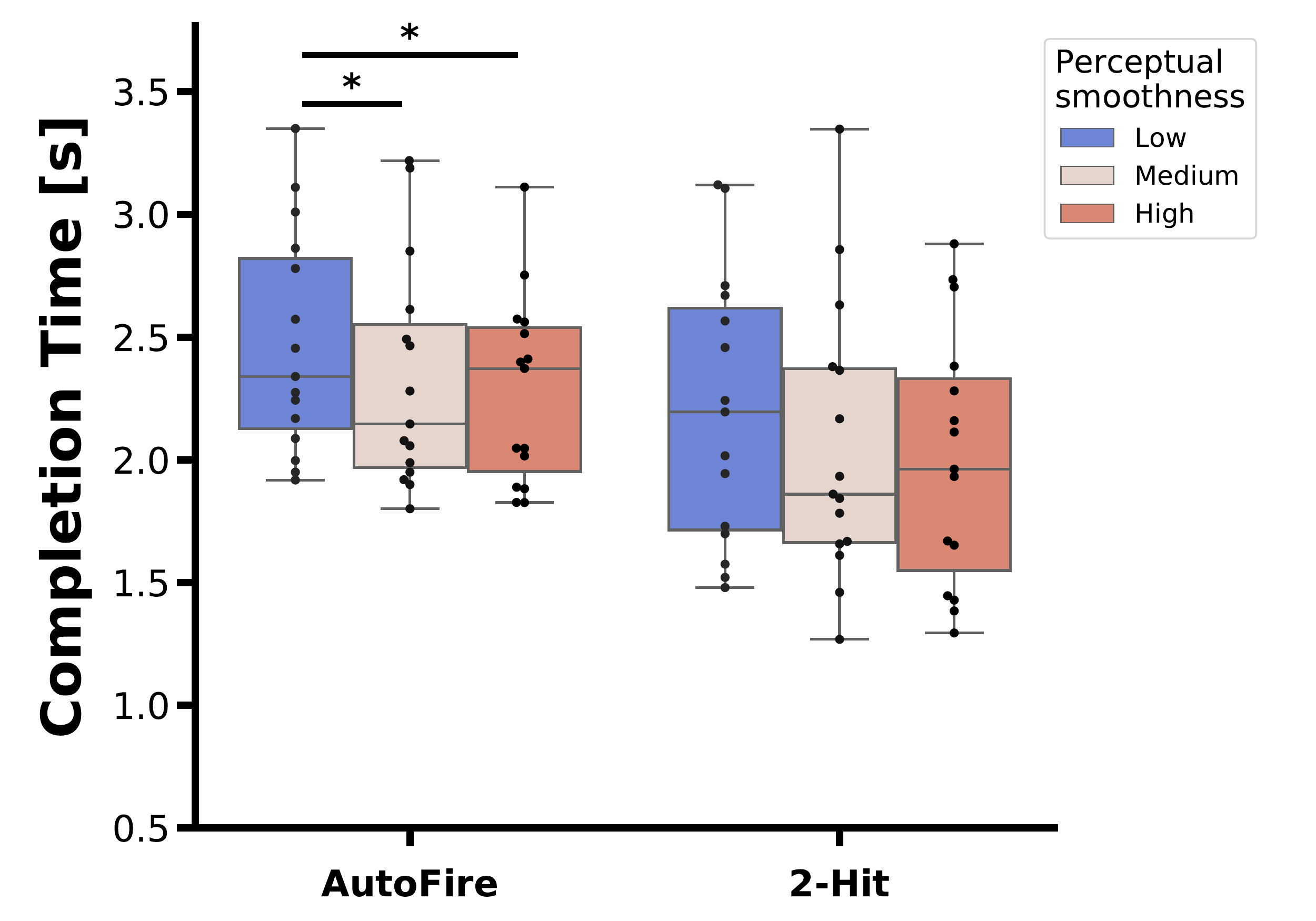}
    \includegraphics[height=52mm]{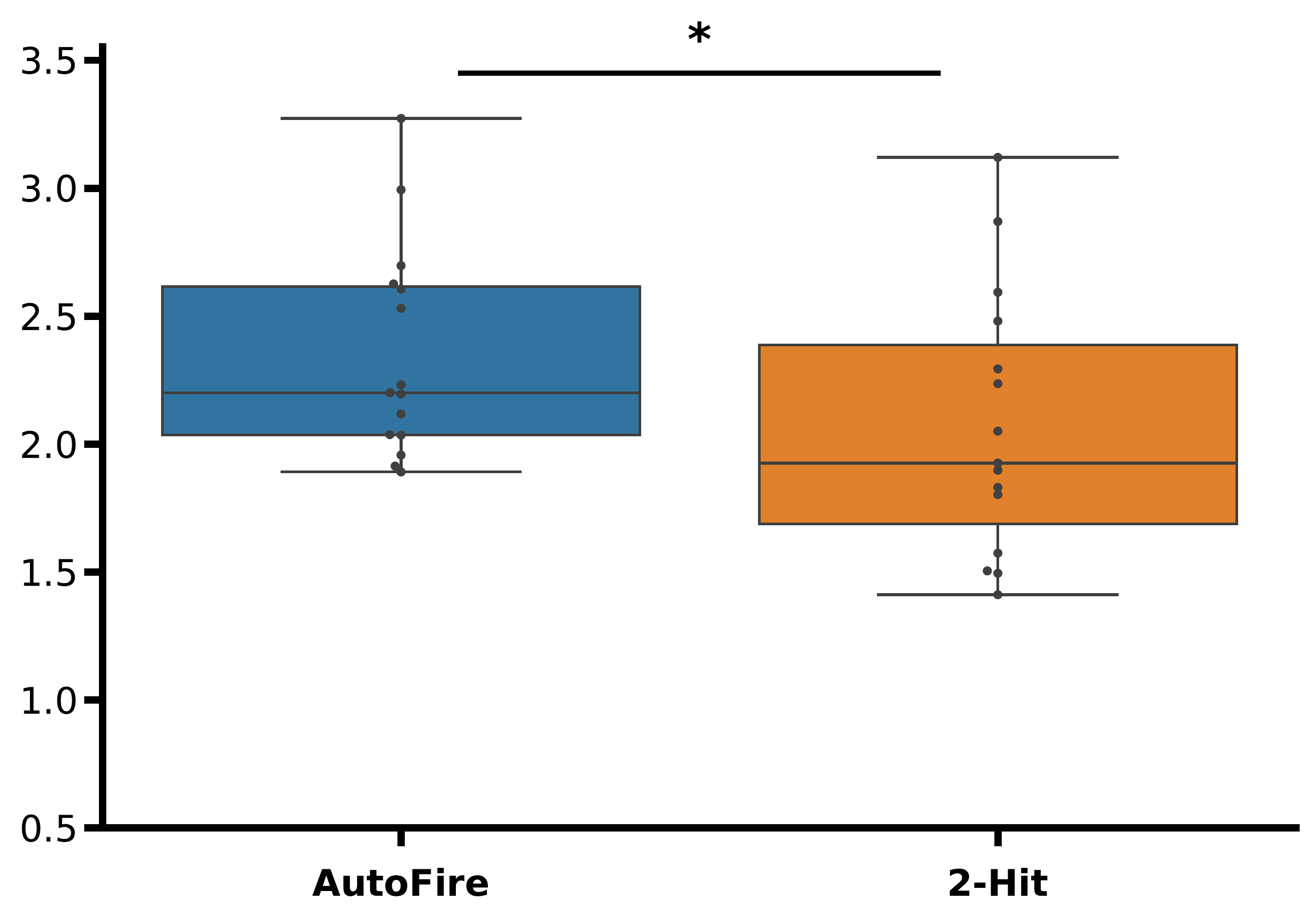}
    \caption{Time to destroy the target. (Left) Box plots with individual data points per subject (n = 15) superimposed on top. The task completion time is stratified by weapon type and perceptual smoothness. (Right) The task completion time is stratified by weapon type only. Horizontal black bars with the * symbol indicate statistically significant comparisons.}
    \label{fig:exp2_TTD}
\end{figure*}

\subsubsection{Time to completion}
The mean time to destroy the target provides us with an overall measure quantifying the impact of variable frame timing on performance in experiment 2.
We observed a significant three-way interaction between weapon configuration, smoothness and variable frame timing, F(2, 28) = 3.490, p = 0.044, ${\eta_{p}^{2}= 0.2}$.
We observed no significant two-way interaction between smoothness and variable frame timing for the Autofire weapon (F(2, 28) = 2.11, p = 0.141, ${\eta_{p}^{2}= 0.131}$), nor a significant main effect for variable frame timing (F(1, 14) = 0.412, p = 0.531, ${\eta_{p}^{2}= 0.029}$), but we did observe a main effect for smoothness (F(2, 28) = 6.28, p = 0.006, ${\eta_{p}^{2}= 0.31}$), as shown in Fig. \ref{fig:exp2_TTD}, left.
Post-hoc analysis between the different levels of smoothness indicated that low smoothness trials took significantly longer to complete (M=2.44 s, SEM=0.104) relative to high smoothness (M=2.28 s, SEM=0.097), t(14) = 3.21, p = 0.006; and average smoothness (M=2.32 s, SEM=0.114), t(14) = 3.01, p = 0.009. 
However, we did not observe a significant difference in time to completion between the medium and high smoothness levels, t(14) = -0.624, p = 0.542.

For the 2-hit weapon configuration, we observed no significant interaction between smoothness and VFT (F(2, 28) = 1.28, p = 0.293, ${\eta_{p}^{2}= 0.084}$) nor a significant main effect for VFT (F(1, 14) = 0.3, p = 0.593, ${\eta_{p}^{2}= 0.021}$).
The main effect of smoothness for the 2-hit weapon was significant at an uncorrected alpha level of 0.05, F(2, 28) = 3.87, p = 0.033, ${\eta_{p}^{2}= 0.216}$, but did not survive a Bonferroni corrected alpha level for the 4 main effects investigated here (alpha = 0.0125). 
Therefore, we did not examine pairwise comparisons between smoothness levels for the 2-hit weapon. 

Our second analysis, which focused on the interaction between the amplitude of the sine wave and weapon configuration conditioned on medium perceptual smoothness, revealed no significant two-way interaction between weapon configuration and amplitude of sine wave, F(3,42) = 2.145, p = 0.109, ${\eta_{p}^{2}= 0.133}$.
We also did not find a significant main effect for amplitude of the sine wave, F(3, 42) = 0.669, p = 0.576, ${\eta_{p}^{2}= 0.046}$ but we observed a significant main effect for weapon configuration, F(1,14) = 47.148, p < 0.001, ${\eta_{p}^{2}= 0.771}$ (Fig. \ref{fig:exp2_TTD}, right).  
Utilization of the 2-hit weapon led to significantly faster task completion time (M=1.94 s, SEM=0.109) relative to the Autofire weapon (M=2.31 s, SEM=0.108).

\section{Discussion}
\label{sec:Discussion}

\subsection{Summary of Results}
In this study, we investigated how VFT influences users' perception of smoothness (experiment 1) and their performance in FPS-style gameplay (experiment 2). 
Our analysis of the results from experiment 1 suggests that VFT impacts our perception of smoothness similarly to CFT (Fig. \ref{fig:exp1_fit}). 
Moreover, as more variation is added to a constant frame time baseline (i.e., larger amplitude), our perception of smoothness becomes tethered to the frames presented on screen for longer periods of time (e.g., the sampled frames near the peak of the sine wave in Fig. \ref{fig:VariableFrameTime}).
This is realized by our MSE analysis (table \ref{tab:percentileMSE}) in the VFT conditions that found higher percentiles of the frequency distributions for each VFT sequence served as better predictors of smoothness than the time-weighted mean frame duration.
For example, we saw a 20\% reduction in MSE when we regressed the smoothness scores of participants in the A = 6 ms condition on the frame duration corresponding to the 70th percentile of each of the five VFT sequences as opposed to the time-weighted mean frame duration. 
However, it should be noted that we fit four free parameters of the sigmoid function to (essentially) five data points (the means smoothness scores across all participants for the VFT sequences).
So some overfitting is to be expected.
Other models (e.g., simple linear regression) may better predict the smoothness in the VFT sequences with considerable variation in frame timing, especially when the VFT sequence oscillates between 1/240 and 1/120 s because it did not intersect the green curve in Fig. \ref{fig:exp1_fit}.
However, this does not preclude our conclusion that frames persisting on screen for the longest period of time in the VFT sequences have a more pronounced influence on the perception of smoothness in gameplay than the frames near the time-weighted mean frame duration of the VFT sequence.

The results from experiment 2 demonstrated that variable frame timing did not significantly impact task completion. 
However, perception of smoothness and the weapon type did account for the variability in task performance. 
Participants took longer to destroy targets with the Autofire weapon in the low smoothness condition relative to the average and high smoothness conditions; however, we did not find the same pattern of results with the 2-hit weapon (Fig. \ref{fig:exp2_TTD} left).
Our second analysis on the effect of the sine wave amplitude, conditioned on average perceptual smoothness, showed no additional interesting effects of VFT and reconfirmed the influence of weapon configuration on completion time.
Taken together, these analyses reject our hypothesis that performance is impacted beyond perceivable changes in VFT.
While frames that last on screen for long periods of time (e.g., 1/30 s) do create significant impacts on task completion time relative to frames that last on screen for short periods of time (e.g., 1/120 s), they do not appear to drive changes in performance amongst CFT and VFT sequences ranked as equally smooth (on average) by users.

\subsection{Latency Effects}
\label{sec:Latency}
\begin{figure}
    \centering
    \includegraphics[width=\columnwidth]{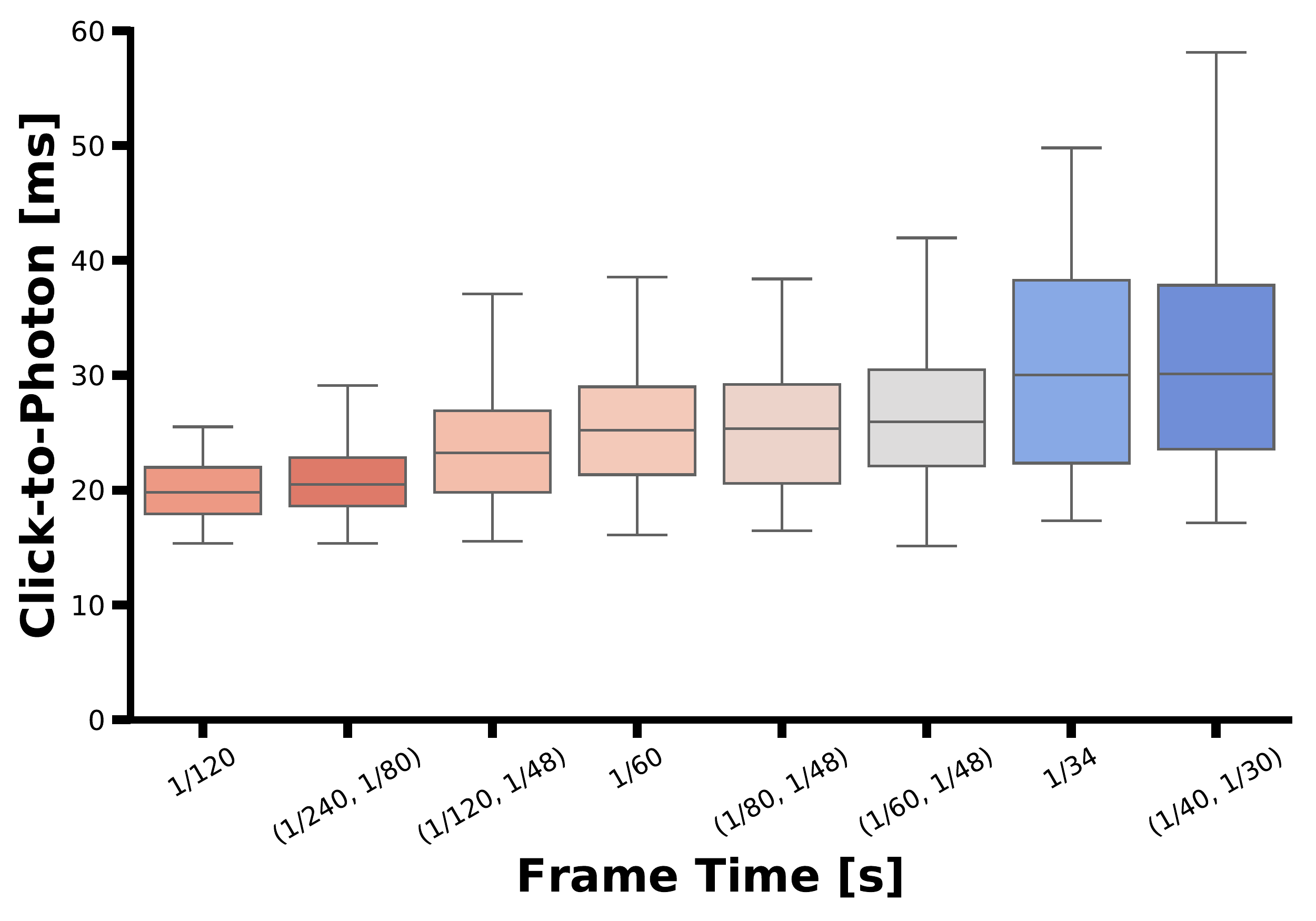}
    \caption{Distribution of click-to-photon latency (measured based on 200 clicks) across the stimuli levels selected for high, mid, and low smoothness for experiment 2 (performance). Horizontal lines indicate median, box bounds upper and lower quartile, and whiskers indicate min and max, excluding outliers. Frame times are sorted by measured latency. Measurements were collected using a click-to-photon tool similar to \cite{schmid2021yet,dossena2022openldat}. Latency may partially explain differences in aiming time (Sec.~\ref{sec:Latency}).}
    \label{fig:LatencyMeas}
\end{figure}

Prior work has demonstrated that end-to-end latency affects FPS aiming task performance~\cite{spjut19latency,ivkovic2015quantifying,liu2021lower} with changes as small as 8 ms showing measurable differences~\cite{spjut2021case}.
Frame time contributes to the latency of a system, with the expected latency linearly proportional to frame duration.
In our experiment, we chose not to attempt to normalize latency across frame time conditions as there was no obvious method for controlling latency during dynamic frame time stimuli.
As a result, different VFT levels result in different mean and range of latencies, as seen in Fig. \ref{fig:LatencyMeas}.
The difference in average latency between the 120 Hz and the 34 Hz constant frame rate conditions was $\sim$10 ms.
Though this likely impacted performance results between the high, mid, and low frame rate conditions, the impact based on the difference in latency between grouped conditions (i.e., 120 Hz fixed vs. (1/240, 1/80) VFT) is likely small.
Future studies on the impacts of VFT on latency and variable latency on performance would be interesting next steps in better understanding these impacts.
We accept that the change in latency across conditions may cause some portion of the measured change in performance in our experiments.

\subsection{Suggestions for Researchers}
Researchers studying frame rate or time as an independent variable in perceptual contexts should carefully consider the effects of variation over time.
Systems used to conduct experiments should have appropriate hardware specifications to guarantee they meet or exceed the desired frame rate and potentially inject additional delay to regularize frame times to as close to constant as possible.
When reporting frame times as a single period, or frame rate, investigators should report variational statistics demonstrating how well controlled their constant frame rate/time was.
When frame rates or times are not sufficiently well controlled, we have demonstrated that they significantly affect the perception of smoothness, with the mean frame time mispredicting the equivalent perception of a constant frame time condition.

We did not observe evidence that, beyond perceptually equivalent conditions, our choice of variable frame timing stimuli impacts FPS aiming task performance.
However, this does not imply that variable frame timing cannot impact task performance.
Specifically, we measured a relatively small frame time variation at typical, lower-end refresh rates.
More considerable variations in frame time, such as the doubling of frames introduced by VSync, could more significantly impact performance through variable frame timing.

\subsection{Suggestions for Gamers}
Gamers should be mindful of additional dynamic workloads and other sources of VFT in their systems.
However, unlike other metrics where performance is often impacted without a perceived difference, variable frame timing appears to be primarily a perceptual effect.
For this reason, gamers are uniquely well suited to evaluate the impacts of variable frame timing on their gameplay as the impacts seem primarily perceptual.
When VFT becomes bothersome to gamers, they have several options for improving the regularity of frame times.
Traditional techniques like buffering larger numbers of frames and enabling VSync can help in some cases, but at the cost of additional latency and increased VFT in certain non-performant cases.
Alternatively, gamers can use VRR technology or approaches like in-game frame rate limiters and adjust settings to reduce additional compute workloads.
Here the primary benefit of VRR is that, unlike more traditional display techniques, it can target constant frame rates other than traditional display refresh rates without exotic display configuration (i.e., 60, 120, 144, 240 Hz).
This allows the gamer to better tailor their ideal, consistent refresh rate to the system capabilities by running at 100 Hz instead of duty cycling between 1/60 s and 1/120 s frames.

\section{Conclusions}
This work focused on the influence of Variable Frame Timing (VFT) on human perception of smoothness and performance in FPS-style gameplay.
It considered a wide range of constant frame time sequences from 1/30 to 1/240 s and variable frame time sequences that oscillated between those typical frame times encountered in the wild.
Across the domain of frame times tested, two user studies demonstrate that large deviations from the average frame time of the VFT sequence affect the perception of smoothness while minimally influencing performance beyond the smoothness effect. 

\begin{acks}
\end{acks}

\bibliographystyle{ACM-Reference-Format}
\bibliography{main}

\end{document}